\documentstyle[newarcrc,fleqn,psfig]{article}

\title{Detached white-dwarf close-binary stars -- CV's extended family}

\author{T.R. Marsh\address{Department of Physics and Astronomy\\
Southampton University, Highfield\\
Southampton SO17 1BJ, England\\
e-mail: trm@astro.soton.ac.uk}}

\begin{document}
\maketitle

\begin{abstract}
I review detached binaries consisting of white dwarfs with either
other white dwarfs or low mass main-sequence stars in tight orbits
around them. Orbital periods have been measured for 15 white
dwarf/white dwarf systems and 22 white dwarf/M dwarf systems.  While
small compared to the number of periods known for CVs ($>300$), I
argue that each variety of detached system has a space density an
order of magnitude higher that of CVs. While theory matches the
observed distribution of orbital periods of the white dwarf/white
dwarf binaries, it predicts white dwarfs of much lower mass than
observed. Amongst both types of binary are clear examples of helium
core white dwarfs, as opposed to the usual CO composition; similar
systems must exist amongst the CVs. White dwarf/M dwarf binaries
suffer from selection effects which diminish the numbers
seen at long and short periods. They are useful for the study of
irradiation; I discuss evidence to suggest that Balmer emission is
broadened by optical depth effects to an extent which limits its
usefulness for imaging the secondary stars in CVs.
\end{abstract}

\begin{table}
\caption[]{Detached white-dwarf/sub-dwarf + white-dwarf/M-dwarf binaries of known orbital period.}
\label{tab:wdbs}
\begin{tabular*}{\hsize}{@{\extracolsep{\fill}}llllp{15mm}llll}
\hline
\multicolumn{4}{c}{WD or sdOB +  WD binaries} & &
\multicolumn{4}{c}{WD or sdOB +  M star binaries} \\
Name          &Period &Type$^a$&Ref$^b$& &Name & Period & Type & Ref  \\	
              &days   &        &     &   &     & days   &      &  \\            
\hline
WD 0957-666   &0.061  & WD/WD  & &    &GD 448        & 0.103 & WD/M   &6 \\
KPD 0422+5421 &0.090  & sdB/WD &1&    &MT Ser        & 0.113 & sdO/M  &  \\
WD 1704+481A  &0.14   & WD/WD  &2&    &HW Vir        & 0.117 & sdB/M & \\  
PG 1101+364   &0.145  & WD/WD  & &    &NN Ser        & 0.130 & WD/M  & \\  
WD 2331+290   &0.166  & WD/WD  & &    &EC 13471-1258 & 0.151 & WD/M  & \\  
PG 1432+159   &0.225  & sdB/?  &3&    &GD 245        & 0.174 & WD/M  & \\  
PG 2345+318   &0.241  & sdB/?  &3&    &BPM 71214     & 0.202 & WD/M  &7 \\  
PG 1101+249   &0.354  & sdB/?  &3, 4& &PG 1224+309   & 0.259 & WD/M  &8 \\  
PG 0101+039   &0.570  & sdB/?  &3&    &AA Dor        & 0.261 & sdO/M &  \\  
WD 1713+332   &1.123  & WD/WD  & &    &CC Cet        & 0.287 & WD/M  &  \\  
WD 1428+373   &1.143  & WD/WD  &2&    &RR Cae        & 0.304 & WD/M  &9 \\ 
WD 1022+050   &1.157  & WD/WD  &5&    &TW Crv        & 0.328 & sdO/M &  \\  
WD 0136+768   &1.407  & WD/WD  &5&    &WD 1042-690   & 0.336 & WD/M  &5 \\  
Feige 55      &1.493  & WD/WD  & &    &GK Vir        & 0.344 & WD/M  &  \\  
L870-2        &1.556  & WD/WD  & &    &KV Vel        & 0.357 & WD/M  &  \\  
WD 1204+450   &1.603  & WD/WD  &5&    &UU Sge        & 0.465 & WD/M  &  \\  
PG 1538+269   &2.50   & sdB/?  & &    &V477 Lyr      & 0.472 & sdO/M &  \\  
WD 1241-010   &3.347  & WD/WD  & &    &Gl 781A       & 0.497 & M/WD  &10 \\  
WD 1317+453   &4.872  & WD/WD  & &    &HZ 9          & 0.564 & WD/M  &  \\  
WD 2032+188   &5.084  & WD/WD  &5&    &PG 1026+002   & 0.597 & WD/M  &11 \\  
WD 1824+040   &6.266  & WD/WD  &5&    &EG UMa        & 0.668 & WD/M  &  \\  
WD 0940+068   &8.33   & sdB/?  &2&    &RE J2013+400  & 0.706 & WD/M  &11 \\  
              &       &        & &    &WD 2009+622   & 0.741 & WD/M  &5 \\  
              &       &        & &    &RE J1016-0520 & 0.789 & WD/M  &11 \\  
              &       &        & &    &IN CMa        & 1.263 & WD/M  &  \\ 
              &       &        & &    &Feige 24      & 4.232 & WD/M  &  \\  
              &       &        & &    &G 203-047 ab  & 14.71 & M/WD  &12 \\
\hline
\multicolumn{9}{l}{\parbox[t]{155mm}{
$^a$ Types defined as primary/secondary with code: WD= white dwarf;
M= M dwarf; sdO/sdB= O/B sub-dwarfs; ?= uncertain.}}\\
\multicolumn{9}{l}{\parbox[t]{155mm}{
$^b$ References are only given if they cannot
be traced from the compilation of \cite{Ritter and Kolb 1998}.}}\\
\multicolumn{9}{l}{\parbox[t]{155mm}{
1.\ \cite{Koen et al 1998}, 
2.\ [Maxted et al., in prep],
3.\ \cite{Moran et al 1999}, 
4.\ \cite{Saffer et al 1998},
5.\ [Moran et al., in prep],
6.\ \cite{Maxted et al 1998},
7.\ [Krzeminski, priv comm], 
8.\ \cite{Orosz et al 1999},
9.\ \cite{Bruch and Diaz 1998}, 
10.\ \cite{Gizis 1998}, 
11.\ \cite{Wood et al 1999}, 
12.\ \cite{Delfosse et al 1999}
}} 
\end{tabular*}
\end{table}

\section{Introduction}
Cataclysmic variable stars (CVs) have not always been as we see them
today. They evolve from pairs of main-sequence stars in relatively
long period orbits. We know this because the white dwarf components of
CVs were once the cores of giant stars much larger than the CVs are
now. The standard explanation for this invokes a phase during which
both stars orbit within a single envelope (derived from the giant
star). As the stars orbit they lose angular momentum to the
envelope which is ejected, leaving a much tighter binary star.

This so-called ``common-envelope phase'' does not produce a CV: some
other angular momentum loss, such as magnetic braking, is required to
further whittle down the orbit before mass transfer from the
still-unevolved secondary star can get underway. Clearly we must expect to
find binary stars which have gone through common-envelope evolution,
but have yet to become CVs. These stars, which for simplicity we will
call pre-CVs -- although they will not always manage to become CVs --
should consist of white dwarf stars with low mass companions, typically
M dwarf stars. I will look at examples of these stars in 
section~\ref{sec:precvs}. They are of direct interest to
evolutionary models of CVs and give us clean examples of irradiated
stellar atmospheres.

After the common-envelope phase, the binary may still be of
sufficiently large separation that it cannot become a CV before the
secondary star has itself evolved. If this occurs one can expect a
second common-envelope phase. If the binary survives this, a pair of
white dwarfs or a ``double-degenerate'' may emerge; such systems may
also be produced from the remnants of Algols. I will refer to them as
DDs. There has been much interest in DDs mainly because they are a
possible progenitor system of Type Ia supernovae. The idea here is
that as gravitational wave radiation shortens their orbital periods,
DDs will eventually start mass transfer at orbital periods of order
100 seconds. While they may survive this (and then emerge as AM~CVn
stars), it is likely instead that they will merge. If the merged
product exceeds the Chandrasekhar limit, collapse will occur which
might ignite fusion violently enough to give a Type Ia supernova, with
no remnant. The biggest problem with this model appears to be whether
explosions occur as opposed to much more gentle collapses leaving
neutron stars; this is largely a matter for theoretical
models. However, a different aspect is directly testable: if DDs are
Type Ia progenitors then there should be a population of DDs with
total masses above the Chandrasekhar limit and with periods short
enough to merge within the lifetime of the Galaxy, which works out at
about 10 hours. I now turn to what is known about DDs.

\section{Double-Degenerates}
The first double-degenerate discovered, L870-2 \cite{Saffer et al
1988}, consists of two cool ($\sim 7$,$000\,{\rm K}$) white dwarfs in
a $1.56$ day period orbit. Around the same time as this discovery,
there were three surveys to find the short period population
relevant to Type Ia supernovae \cite{Robinson and Shafter 1987,
Bragaglia et al 1990, Foss et al 1991}. These were mostly
unsuccessful, although a system called 0957-666 was found to have a
1.18 day period \cite{Bragaglia et al 1990}.

Soon after this work, model atmosphere and evolutionary model fits to
the spectra of white dwarfs revealed a population of low mass
($<0.45\,{\rm M}_\odot$) objects \cite{Bergeron et al 1992, Bragaglia
et al 1995}. On the other hand, white dwarfs which evolve from
single stars within the age of the Galaxy are expected to have a 
minimum mass of around $0.55\,{\rm M}_\odot$. The models are dependent
upon the uncertainties of mass loss on the AGB, but some white dwarfs
have masses as low as $0.33\,{\rm M}_\odot$, which is too low for them
even to have reached the AGB. These must be the helium cores of stars
which failed to advance beyond the RGB, perhaps as a result of
mass loss within a binary. This suggested that concentrating on the low
mass white dwarfs might be an effective method for finding close
binaries, as indeed proved to be the case \cite{Marsh et al 1995,
Marsh 1995}. This has raised the number of DDs with measured periods
to 15, with another 7 sdB binaries that probably have white dwarf
companions (see table~\ref{tab:wdbs}). During this work it was also
found that the original period determination for 0957-666 was in
error; the revised value of $0.061$ days remains the shortest
known for these systems \cite{Moran et al 1997}.

\begin{figure}[htb]
\begin{minipage}[t]{77mm}
\raisebox{-3mm}{\parbox[t]{75mm}{\psfig{file=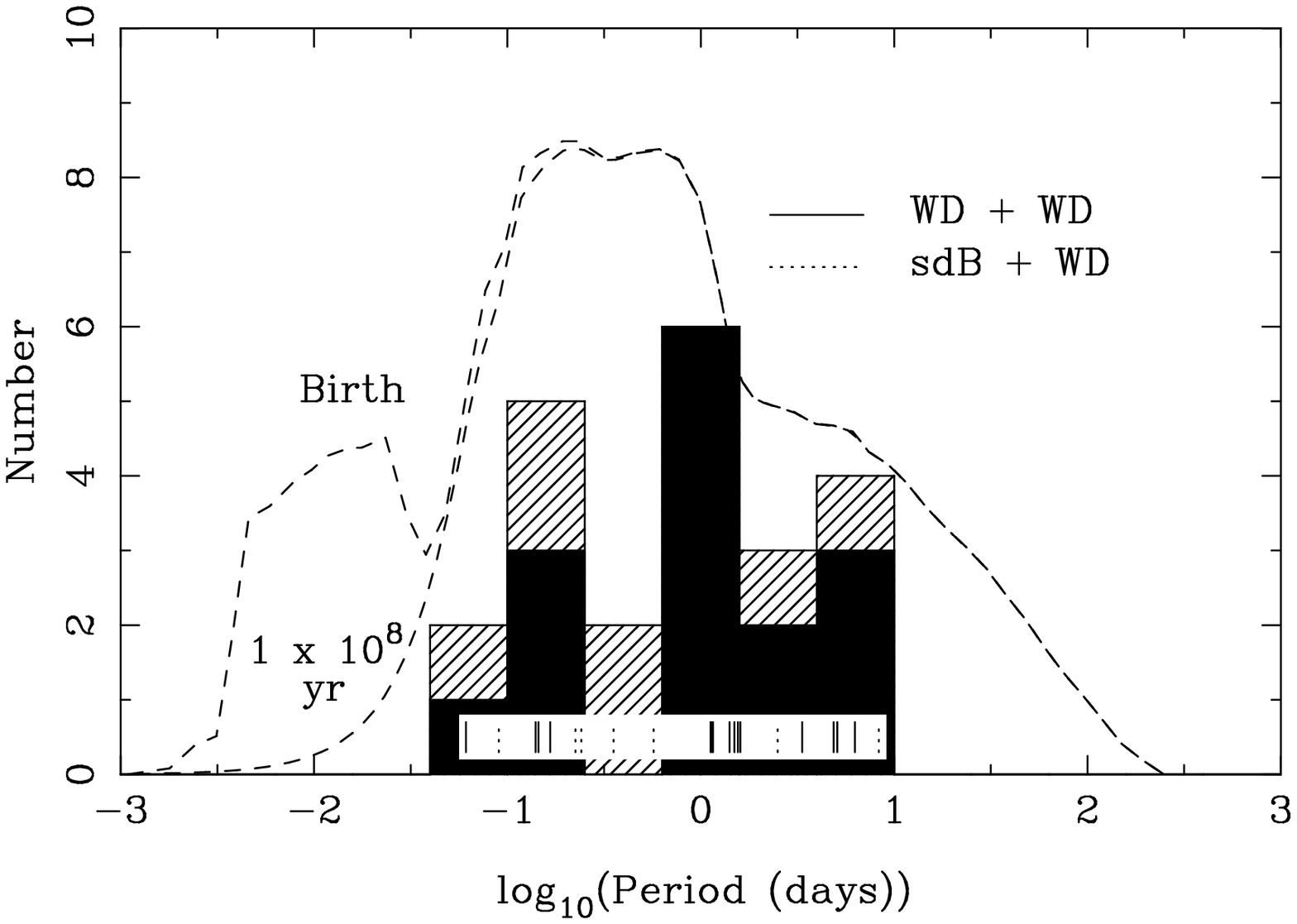,width=75mm}}}
\caption{The period distribution of WD+WD and
sdB+WD binaries is compared to theoretical distribution at birth
and after $10^8$ yr of erosion by gravitational waves 
\cite{Iben et al 1997}. (Solid = WD+WD only; hatched includes sdB+WD
too.)}
\label{fig:dd_pdist}
\end{minipage}
\hspace{\fill}
\begin{minipage}[t]{77mm}
\psfig{file=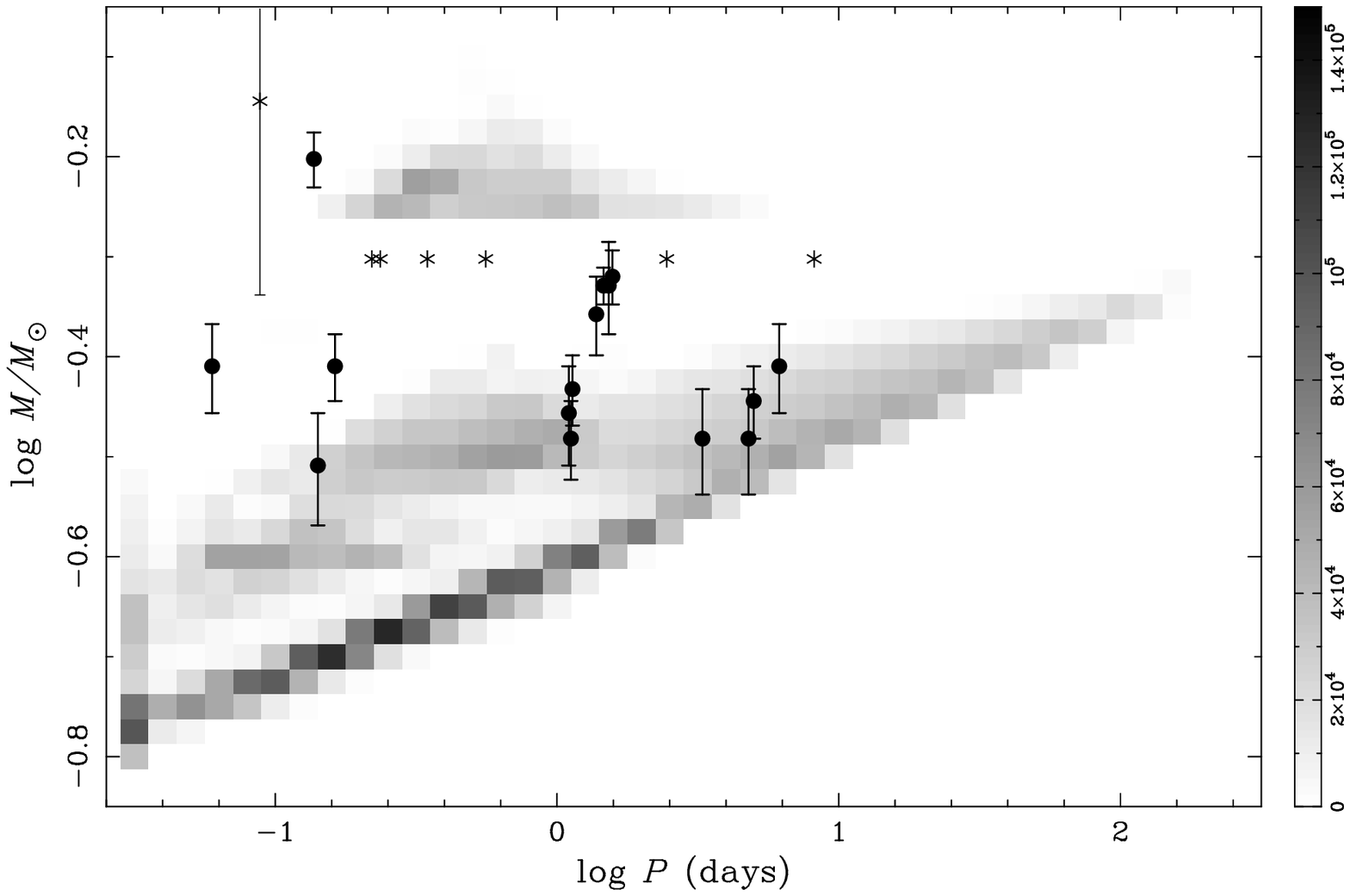,width=75mm}
\caption{The mass versus period distribution of WD+WD (solid circles)
and sdB+WD (asterisks) binaries is compared to theory. 
Update of \cite{Saffer et al 1998}.}
\label{fig:saffer}
\end{minipage}
\end{figure}

The observed periods are compared to the results of binary
``population synthesis'' \cite{Iben et al 1997}
in Fig.~\ref{fig:dd_pdist}. I have assumed that all the sdB stars
in the left column of Table~\ref{tab:wdbs} have white dwarf companions and
that they will emerge as DDs with little alteration in period; this
remains to be proved. The essential result of the comparison is that
theory and observation match fairly well, although there is perhaps 
a hint that there may be a dearth of DDs with periods around $0.5$ 
days. 

Things become more interesting when one looks at the 
2-dimensional distribution of mass and period (Fig.~\ref{fig:saffer});
the relative reliability of mass determination for non-accreting
white dwarfs is a significant advantage compared to the normal case
for CVs. Fig~\ref{fig:saffer} shows a significant discrepancy between
theory and observation. Theory predicts the existence of a large
fraction of very low mass white dwarfs ($\sim 0.25\,{\rm M}_\odot$)
which are not observed; I can think of no plausible observational
selection effect to side-step this discrepancy. Reinforcing this
problem, particular systems, such as 0957-666 (the left-most point),
lie in regions of near-zero probability according to theory. While
the theory has many free parameters that can be adjusted to produce
a better fit, the absence of very low mass white dwarfs is a puzzle
as it suggests that for some reason we never see the results of mass
loss early on the RGB. 

\subsection{Numbers of DDs}
With only 15 bona-fide DDs with measured periods, compared to over 300
CVs \cite{Ritter and Kolb 1998}, it may seem that they are relatively
rare. In fact the reverse is the case: my best guess at the space
density of DDs is $5 \times 10^{-4}\,{\rm pc}^{-3}$, of order 20 times
that of CVs, including the very faint and so far undetected CVs
presumed to have ``bounced'' at 80 mins orbital period \cite{Politano
1996}. The estimate for DDs is based on the relatively well determined
space density for all white dwarfs \cite{Knox et al 1999} and the
roughly 10\% of white dwarfs that are DDs \cite{Saffer et al 1998,
Maxted and Marsh 1999}. The difference in observed numbers is down to
ease of detection. This means that there are some 250 million DDs in
the Galaxy, with perhaps 1 million systems with periods of less than
an hour; they are likely to be the dominant source of low frequency
gravitational waves in the Galaxy \cite{Hils et al 1990}.

Can DDs be the progenitors of Type Ia supernovae? We have now found
systems of short enough period, and one, KPD 0422+5421 \cite{Koen et
al 1998}, may even have enough mass. In terms of numbers, and leaving
aside the issue of whether they really explode on merging, the answer
would appear to be yes, they remain a viable progenitor. While we have
not found convincing examples of systems with enough mass, these are
probably just rare; only about 1 in 40 of DDs is expected to be such a
system \cite{Iben et al 1997} and we have been concentrating
specifically on low mass systems.

\section{Pre-CVs}
\label{sec:precvs}
When one searches for DDs, one also finds pre-CVs. I define these as
binaries containing a white dwarf (or sub-dwarf which will evolve into
a white dwarf) and an M dwarf companion. Higher mass companions are
excluded because (a) it becomes hard to see the white dwarf if the
companion is too bright and (b) theoretically, CVs are descended from
systems with mass ratio $q = M_{\rm MS}/M_{\rm WD} < 0.28$
\cite{Politano 1996}, and since white dwarfs are usually below a solar
mass, this implies M dwarf companions. There are 27 pre-CVs known; 22
are white dwarf/M dwarf systems, and 5 are sub-dwarf/M dwarf systems.

\begin{figure}[htb]
\begin{minipage}[t]{77mm}
\psfig{file=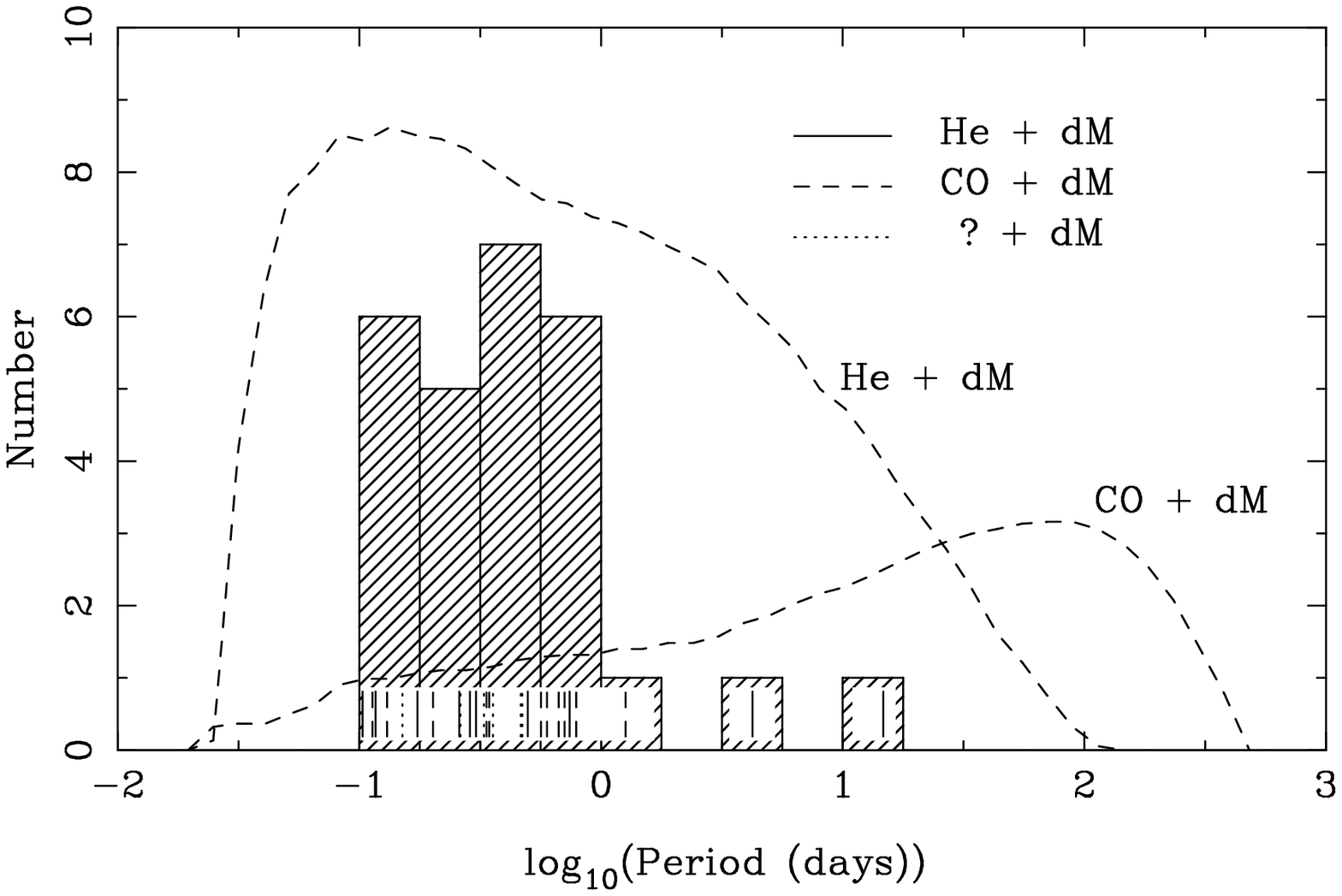,width=75mm}
\caption{The period distribution of white dwarf or sub-dwarf + M dwarf
binaries is compared to theory \cite{Iben et al 1997}. The systems
have been classified into helium or CO primaries according to whether
the mass is greater than $0.49\,{\rm M}_\odot$ or not; for some
systems the mass is unknown.}
\label{fig:precv_pdist}
\end{minipage}
\hspace{\fill}
\begin{minipage}[t]{77mm}
\raisebox{5mm}{\parbox[t]{75mm}{
\psfig{file=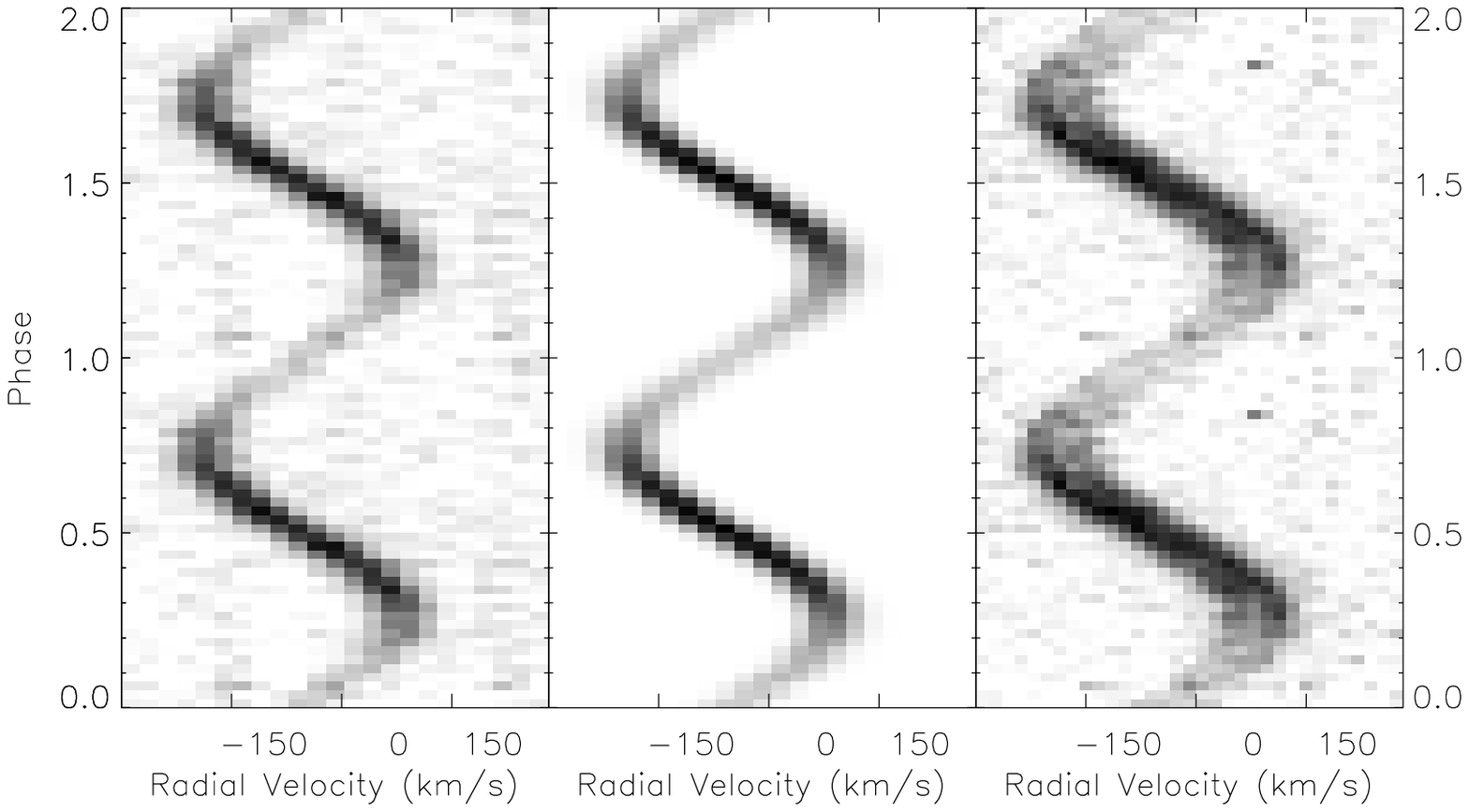,width=75mm}}}
\caption{Trailed spectra of the $2.5$ hour white dwarf/M dwarf binary
GD~448 show irradiation-induced emission from the M dwarf. The
H$\alpha$ emission (right panel) comes from the same place as the CaII
(left panel) but is broadened by optical depth effects \cite{Maxted et
al 1998}. The centre panel shows a model with no broadening other than
instrumental.}
\label{fig:gd448}
\end{minipage}
\end{figure}

The observed periods are compared to theory in Fig.~\ref{fig:precv_pdist}.
Observations and theory do not compare well. In this case however, I
think it is likely that observational selection could be to blame for
the lack of systems at both long and short periods. At long periods,
the radial velocity of the white dwarf is relatively low and
irradiation-induced emission from the M dwarf will be weak. There are
a good number of white dwarf/M dwarf pairs known which don't have
measured orbital periods, and the long period systems may well be
lurking amongst them. At short periods only very low mass M stars can
remain inside their Roche lobe; for example the shortest period system
listed, GD~448, has a mass of $0.09\,{\rm M}_\odot$, barely above the
brown dwarf limit \cite{Maxted et al 1998}. It is difficult to see any
sign of the M dwarf in GD~448, with only weak emission at H$\alpha$;
we may well be missing still shorter period systems.

The masses of the white dwarfs in pre-CVs are not as well determined
as they are for DDs because the line profiles are often filled in by
emission from the M dwarf. However, there are enough known to be
certain that helium core white dwarfs exist in some numbers. I define
helium-core white dwarfs as those with masses $< 0.5\,{\rm M}_\odot$;
some are borderline, but there is little doubt for systems such as
GD~448 ($M_{\rm WD} = 0.41\pm0.01\,{\rm M}_\odot$) and RR~Cae
($0.36\pm0.04\,{\rm M}_\odot$).  Therefore there must be helium-core
white dwarfs amongst the CVs too, as expected theoretically
\cite{Politano 1996}, although observational selection effects which
favour high masses may count against their detection.

There are four eclipsing white dwarf/M dwarf systems known (GK~Vir,
RR~Cae, NN~Ser and EC 1347-1258). Observations of these have the
potential to provide accurate system parameters and to detect orbital
period changes as may be caused by solar-type magnetic cycles. These
systems cover a range of M dwarf mass, and it would be particularly
interesting, for example, to see if the period of RR~Cae, which has 
a very low mass M dwarf ($0.09\,{\rm M}_\odot$), changes since the
standard disrupted magnetic braking model would suggest a low level of
magnetic activity in such a star.

The numbers of pre-CVs are comparable to DDs i.e.\ they are
intrinsically much more common than CVs. It may prove difficult
to detect potential period-gap-crossing systems against this
background.

\subsection{Irradiation in pre-CVs}
The pre-CVs provide clean systems for the study of irradiation of 
stars. A result of interest for CV studies is that the Balmer
emission lines induced by irradiation are significantly broadened
by optical depth effects (Fig.~\ref{fig:gd448}), \cite{Maxted et al
1998, Wood et al 1999}. The broadening is of order $40\, {\rm
km}\,{\rm s}^{-1}$, which is enough to severely limit their usefulness
for imaging the secondary star in CVs; the CaII lines seem to be a
better option (Fig.~\ref{fig:gd448}). The same broadening is seen in
chromospherically active stars, which is perhaps surprising given
the rather different mechanisms producing the lines.

\section{Conclusions}
Over the last ten years the number of double-degenerate binaries has
gone from 1 to 15 and it is apparent that they are intrinsically
extremely common within our Galaxy, with a space density of order
$5 \times 10^{-4}\,{\rm pc}^{-3}$. Their short periods, which range
from $1.5$ hours to a few days, are a testament to the orbital
shrinkage involved in ejecting the envelopes of the two white dwarfs.
In terms of numbers, they remain a viable progenitor class for Type Ia
supernovae.

The pre-CVs have grown similarly in number, although observational
selection affects detection at periods above a day or so and
below two hours. Amongst them are helium core white dwarfs, and
presumably this must be the case for CVs too. Irradiation-induced
Balmer emission is broadened by radiative transfer effects, and should
be avoided in favour of CaII for imaging the secondary stars
in CVs.

\section{Acknowledgements}
I thank Pierre Maxted for many conversations about these systems, and
Zhanwen Han, Lev Yungelson and Gijs Nelemans for insights into 
evolutionary theory.

\end{document}